\documentclass[lettersize,journal]{IEEEtran}
\usepackage{amsmath,amsfonts}

\usepackage{array}
\usepackage{textcomp}
\usepackage{stfloats}
\usepackage{url}
\usepackage{verbatim}
\usepackage{graphicx}
\usepackage{cite}
\usepackage{subfigure}
\usepackage{subcaption}
\usepackage{adjustbox} 
\usepackage{array}     
\usepackage{multirow}
\usepackage{booktabs}

\usepackage{amssymb}
\usepackage{xcolor}

\begin{document}

\title{Instance Communication System for Intelligent Connected Vehicles: Bridging the Gap from Semantic to Instance-Level Transmission}

\author{Daiqi Zhang, Bizhu Wang, Wenqi Zhang, Chen Sun, IEEE Senior Member,  Xiaodong Xu, IEEE Senior Member
\thanks{This work has been submitted to the IEEE for possible publication. Copyright may be transferred without notice, after which this version may no longer be accessible. This work is supported by the National Natural Science Foundation of China No. 62501067 and Sony (China) Research Funding (Corresponding author: Bizhu Wang).}%
\thanks{Daiqi Zhang, Bizhu Wang and Xiaodong Xu are with the State Key Laboratory of Networking and Switching Technology, Beijing University of Posts and Telecommunications, Beijing, China. Wenqi Zhang and Chen Sun are with Sony (China) Research Laboratory.}%
}



\maketitle

\begin{abstract}
Intelligent Connected Vehicles (ICVs) rely on high-speed data transmission for efficient and safety-critical services. However, the scarcity of wireless resources limits the capabilities of ICVs. Semantic Communication (SemCom) systems can alleviate this issue by extracting and transmitting task-relevant information, termed semantic information, instead of the entire raw data. Despite this, we reveal that residual redundancy persists within SemCom systems, where not all instances under the same semantic category are equally critical for downstream tasks. To tackle this issue, we introduce Instance Communication (InsCom), which elevates communication from the semantic level to the instance level for ICVs. Specifically, InsCom uses a scene graph generation model to identify all image instances and analyze their inter-relationships, thus distinguishing between semantically identical instances. Additionally, it applies user-configurable, task-critical criteria based on subject semantics and relation-object pairs to filter recognized instances. Consequently, by transmitting only task-critical instances, InsCom significantly reduces data redundancy, substantially enhancing transmission efficiency within limited wireless resources. Evaluations across various datasets and wireless channel conditions show that InsCom achieves a data volume reduction of over 7.82 times and a quality improvement ranging from 1.75 to 14.03 dB compared to the state-of-the-art SemCom systems.
\end{abstract}


\section{Introduction}
\par



Intelligent connected vehicles (ICVs) rely on AI services such as collision warning to enhance transportation efficiency and safety.\cite{Wan2024JIOT_TargetDetect,GimenezGuzman2024MNET_SemV2X} However, these services require real-time transmission of large amounts of image and video data, which conflicts with the limited wireless spectrum resources\cite{GimenezGuzman2024MNET_SemV2X,Wan2024JIOT_TargetDetect}. Semantic communication (SemCom) offers a new approach to this challenge\cite{Xu2023MCOM_DeepJSCC_SemCom,Huang2025JSAC_D2JSCC}. Rather than transmitting data completely in bits, it focuses on extracting and conveying the meaning of the data. For example, after a leading vehicle captures a traffic image, only the semantic information extracted by a semantic encoder needs to be transmitted\cite{GimenezGuzman2024MNET_SemV2X,Eldeeb2024LWC_MultiTaskAV}. The following vehicle can then decode this information to reconstruct the image, identify potential hazards, and initiate timely braking\cite{Eldeeb2024LWC_MultiTaskAV,Wan2024JIOT_TargetDetect}. Since the volume of semantic information is much smaller than that of raw data, integrating SemCom into ICVs enables faster and more efficient exchange of critical information between vehicles.\par

While existing SemCom systems offer improvements over traditional approaches, opportunities remain for further data reduction to mitigate the tension between the high-rate data transmission demands and the limited wireless resources in ICV scenarios\cite{Im2024JIOT_AttentionAware,Choi2025LCOMM_ROIJSCC}. Taking the state-of-the-art (SOTA) SemCom system NTSCC as an example, it enhances joint source-channel semantic coding by employing a variable-rate mechanism\cite{Dai2022JSAC_NTSCC,Wang2023_NTSCCpp}. This integration improves the robustness and spectral efficiency in transmitting critical semantic information in time-varying channels\cite{Wang2023_NTSCCpp,Gao2024TCCN_SemAMR}. For example, in ICV scenarios, NTSCC prioritizes safety-critical semantics, such as the pedestrian, over background elements like the sky\cite{Im2024JIOT_AttentionAware,Choi2025LCOMM_ROIJSCC}. Consequently, extracted semantic features of pedestrians (including those walking on the sidewalk and street) are allocated more wireless resources for transmission than those of background areas. However, for tasks like pedestrian collision warning, only specific instances, such as pedestrians walking on the street, are truly critical\cite{Choi2025LCOMM_ROIJSCC}. In conclusion, current SemCom systems lack task-aware instance discrimination, leading to residual redundancy.\par
 
To address this gap, we propose an Instance Communication (InsCom) system that elevates transmission from the semantic level to the instance level. The key contributions are:\begin{itemize}
    \item The proposed InsCom system represents the first instance-level transmission framework for ICVs, advancing beyond SemCom systems by enabling reliable and efficient transmission of task-critical instances within the constraints of limited wireless resources.
    
    \item The proposed InsCom system employs a scene graph generation model to identify all image instances and analyze their inter-relationships, enabling differentiation among distinct instances sharing identical semantics. Furthermore, it employs configurable task-critical criteria based on subject semantics and relation-object pairs to filter recognized instances, transmitting only task-critical instances while significantly reducing data redundancy.
    
    \item Evaluations across various datasets and wireless channel conditions demonstrate that InsCom achieves a data volume reduction of over 7.82 times and a quality improvement ranging from 1.75 to 14.03 dB compared to the SOTA SemCom systems.
\end{itemize}

\section{Proposed InsCom System}

\begin{figure*}[t]
\centering
\includegraphics[width=0.8 \linewidth]{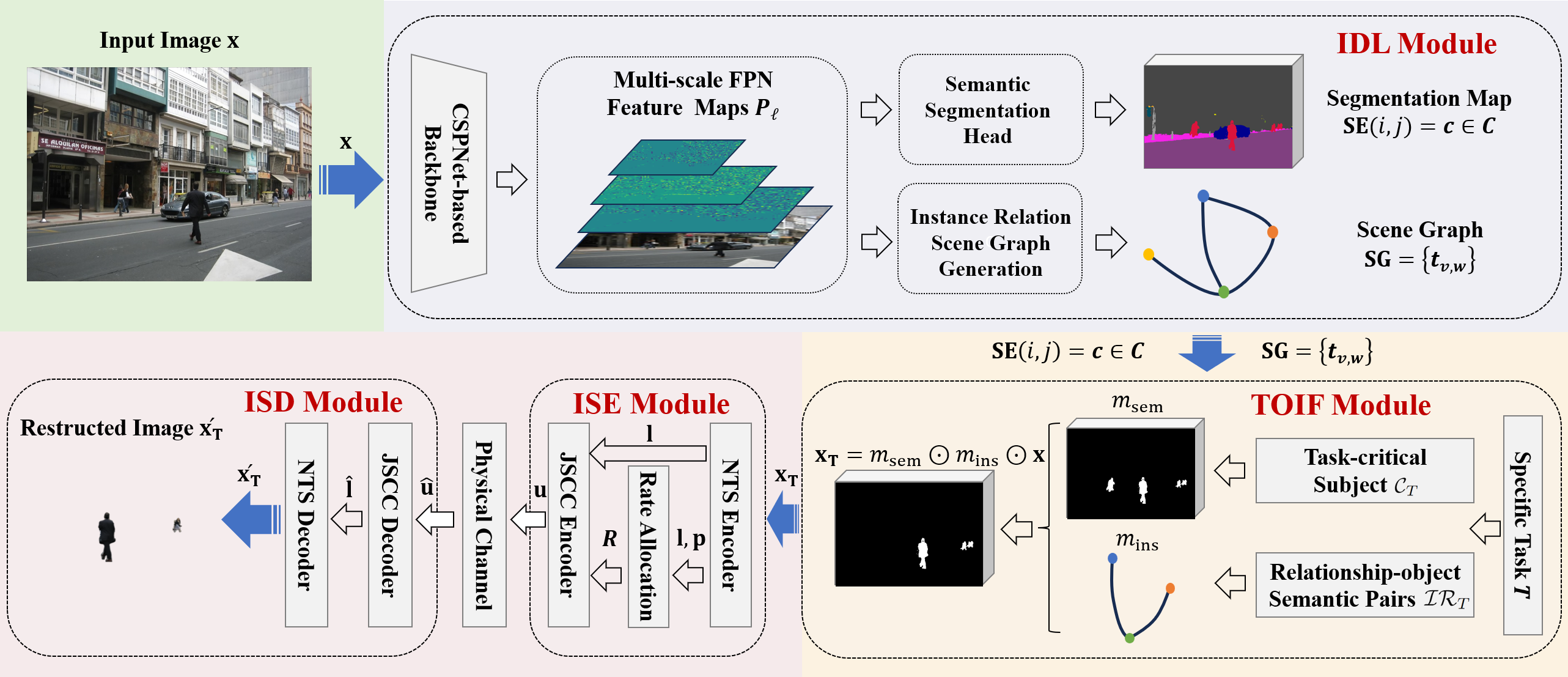}
\caption{The model structure of the proposed InsCom System.}
\label{system_model}
\vspace{-0.5cm}
\end{figure*}

As illustrated in Fig.~\ref{system_model}, InsCom introduces an instance-centric pipeline consisting of four core modules: Instance Differentiation and Localization (IDL), Task-Oriented Instance Filtering (TOIF), Instance Semantic Encoding (ISE), and Instance Semantic Decoding (ISD). The roles of these modules are summarized as follows:
\begin{itemize}
    \item IDL Module: Performs instance-level parsing on the source image $\mathbf{x}$, outputting an instance segmentation map $\mathbf{SE}$ and a structured scene graph $\mathbf{SG}$ that represents subject--object instances and their relationships, enabling precise differentiation and localization of semantically identical instances.
    \item TOIF Module: Configures task-critical criteria via user-defined subject semantics and relation--object pairs, filters instances in $\mathbf{SG}$ based on these criteria, generates a mask $\mathbf{m_T}$ from the bounding boxes of the retained instances and $\mathbf{SE}$, and outputs the masked image $\mathbf{x_T} = \mathbf{x} \odot \mathbf{m_T}$ to preserve critical instances and reduce redundancy.
    \item ISE Module: Extracts semantic features from $\mathbf{x_T}$ and maps them to channel symbols using variable-rate coding, thereby optimizing transmission performance under wireless resource constraints.
    \item ISD Module: Reconstructs $\mathbf{x'_T}$ from the received channel symbols by reversing the operations of ISE, ensuring the reliable recovery of task-critical instances.
\end{itemize}

\subsection{Instance Differentiation and Localization Module}
\label{subsec:idlm}
As shown in Fig.~\ref{system_model}, the IDL module carries out a detailed instance-level semantic parsing on the source image $\mathbf{x}$. This module identifies all instances within the image, delineates their precise contours, and analyzes the relationships between instances. Consequently, the IDL module produces a structured scene graph $\mathbf{SG}$ and an instance segmentation map $\mathbf{SE}$. By enabling the differentiation of instances across varying semantics as well as among distinct instances within the same semantic category, the IDL module lays a solid foundation for further configurable and task-driven instance filtering processes.\par


The IDL module first processes the source image $\mathbf{x}$ through a CSPNet-based backbone and a PAN-FAN-based neck network, generating multi-scale feature maps enriched with semantic and positional cues for object detection. These features are then further processed by a scene graph generation head and a segmentation head. The procedure is outlined below.\par

\subsubsection{Scene Graph Generation Head}
The IDL module utilizes the head network of YOLOv8m, which operates on each multi-scale feature maps from the neck network. This head predicts the probability of containing an instance, the instance semantic-class probabilities, and the bounding box offsets for each grid cell. For each detected bounding box, the pixels within its region are fed into a RoIAlign module to extract visual features from the backbone feature maps. Thus, the set of instance representations for image $\mathbf{x}$ can be defined as
\begin{equation}
    \mathcal{I} = \{i_k\}_{k=1}^K = \{(f_k, c_k, s_k, b_k)\}_{k=1}^K
\end{equation}
where \(i_k\) denotes the representation of the \(k\)-th instance in \(\mathbf{x}\), with \(\mathbf{f}_k\) as the visual feature, \(c_k\) as the semantic class label, \(s_k \) as the confidence score, and \(\mathbf{b}_k\) as the normalized bounding box coordinates. \(K\) is the total number of instances in \(\mathbf{x}\).\par

For any pair of instances \(v\) and \(w\), the semantic fusion module constructs a joint feature representation \(I_{v,w} = (i_v, i_w)\). The relation decoder subsequently refines semantic label predictions for both instances while predicting their relationship. We repurpose the symbol $c$ to denote the optimized semantic labels. Optionally, the causal intervention module refines the relationship prediction to \(y_{v,w}\), thereby reducing biases.\par
This process consequently generates a structured triplet for each instance pair, denoted as 
\begin{equation}
    t_{v,w} = \langle subject, relation, object \rangle=   \langle s_v, r_{v,w}, o_w \rangle 
\end{equation}

Here, \( s_v \) denotes the subject instance with a semantic label \( \phi(s_v)=c_v \), \( o_w \) represents the object instance with a semantic label \( \phi(o_w)=c_w \)  and \( r \) defines the relationship predicate between them, with a semantic label \( \phi(r_{v,w})=y_{v,w} \).\par
Thus, the complete scene graph from $\mathbf{x}$ is defined as:
\begin{equation}
    \mathbf{SG} = \{t_{v,w} \mid v \neq w, v, w = 1, \ldots, K\}
\end{equation}

In summary, the scene graph generation head in the IDL module leverages relational context to distinguish instances with identical semantic labels (e.g., a woman walking on the street versus a woman walking on the sidewalk). This capability forms the basis for the subsequent task-oriented instance filtering process, which is detailed in Section~\ref{subsec:toifm}.

\subsubsection{Segmentation Head}
Since the bounding boxes \(\mathbf{b_k}\) identified by the scene graph generation head are rectangles that inevitably include not only the target instance's pixels but also those of other instances or the background, extracting and transmitting features from all pixels within these boxes leads to redundancy. To address this, IDLM incorporates a Segmentation Head to produce pixel-wise segmentation masks to reduce the transmitted data volume.\par


The segmentation head employs an encoder-decoder design to hierarchically fuse multi-level features. High-level features extract multi-scale context, while low-level features preserve structural details. After concatenation, a lightweight prediction layer performs pixel-wise classification, producing a low-resolution map that is upsampled to the input resolution $H \times W$. This process results in a high-resolution semantic segmentation map $\mathbf{SE}\in \mathbb{Z}^{H \times W}$, defined as:
\begin{equation} 
\mathbf{SE}(i, j) = c \in \mathcal{C}
\end{equation}

Here, $i \in [0, H-1]$ and $j \in [0, W-1]$ represent the row and column indices of the pixel, respectively, and $\mathbf{SE}(i, j)$ denotes the pixel value at position $(i, j)$, which corresponds to the semantic class label $c$.

\subsection{Task-Oriented Instance Filtering Module}
\label{subsec:toifm}
The TOIF Module tailors task-critical information for diverse ICV tasks by leveraging configurable criteria derived from subject semantics and relation-object pairs. It then performs two-stage filtering on $\mathbf{SG}$ using these criteria, generates mask $\mathbf{m_T}$ from retained instances' bounding boxes and $\mathbf{SE}$, and outputs the masked image $\mathbf{x_T}$. This process minimizes data transmission to subsequent ISE modules while mitigating resource expenditure on non-critical semantics and task-irrelevant instances. 

\subsubsection{Task-critical Criteria Configuration}
The TOIF module's configurable criteria consist of:
\begin{itemize}
    \item $\mathcal{C}_{T}$: Subject semantics critical to task $T$ \\ (e.g., \textit{vehicle}, \textit{pedestrian})
    \item $\mathcal{IR}_{T}$: Relationship-object pairs critical to task $T$ \\ (e.g., $\langle \text{walking on}, \text{street} \rangle$, $\langle \text{on}, \text{crosswalk} \rangle$)
\end{itemize}

This parameterization permits efficient task-specific adaptation in ICV deployments.

\subsubsection{Two-stage Instance Filtering}
The TOIF module employs a two-stage instance filtering mechanism, customized based on the aforementioned task-critical criteria. The procedure is detailed below:
\begin{itemize}
    \item Semantic-level Filtering: TOIF processes $\mathbf{SG}= \{t_{v,w}\}=\{\langle s_v, r_{v,w}, o_w \rangle\}$ to retain only triplets where the subject's semantic label $\phi(s_v)=c_v$ belongs to $\mathcal{C}_{T}$, generating the intermediate subgraph $\mathbf{SG}^{1}$. Taking the pedestrian collision warning task as an example, this retains all triplets with `man' and `woman' as subject (e.g., $\langle$ woman, walking on, street $\rangle$ and $\langle$ woman walking on sidewalk $\rangle$, while discarding triples with irrelevant subject instances like buildings or vegetation.
    \item Instance-level Filtering: The graph $\mathbf{SG}^{1}$ is refined by retaining only the triplets that satisfy the condition $\langle r_{v,w}, \phi(o_w) \rangle \in \mathcal{IR}_{T}$. This process results in the final task-critical subgraph $\mathbf{SG}^{2}$. For instance, although both subject instances labeled as “woman” are considered, only the triplet $\langle$woman, walking on, street$\rangle$ is retained, while the triplet $\langle$woman, walking on, sidewalk$\rangle$ is discarded.
\end{itemize}

\subsubsection{Masked Image Generation}

For each task-critical semantic $ c \in \mathcal{C}_{T} $, the corresponding segmentation mask $ m_{\text{sem}} $, which retains the pixels in $ \mathbf{SE} $ whose semantic labels belong to the task-critical semantics set $ \mathcal{C}_{T} $, is denoted as:

\begin{equation}
m_{\mathrm{sem}} = \mathbf{1}\big[(i,j)|S(i,j) \in \mathcal{C}_T\big].
\label{eq:msem}
\end{equation}

Here $i$ and $j$ denote coordinates of the pixels. 
Then, each subject instance within $ \mathbf{SG}^{2} $ is named as a task-critical instance, and the set of task-critical instances is denoted as $S_{T} = \{s_{t}\} = \{s_{v} \mid \langle s_{v}, r_{v,w},o_w \rangle \in \mathbf{SG}^{2}\} $. For each $ s_{t} $, the bounding box is $ \mathbf{b}_{t} $, and the corresponding set is denoted as $ B_{T} = \{\mathbf{b}_{t}\}$. Then, the instance-level mask $ m_{\text{ins}} $ is defined by:

\begin{equation}
m_{\mathrm{ins}} = \mathbf{1}\big[(i,j) \in B_{T}\big].
\label{eq:msem}
\end{equation}

The final mask $ m_{\mathrm{T}} $ is the intersection region of $ m_{\text{sem}} $ and $ m_{\text{ins}} $, and the task-oriented masked image is formed as:

\begin{equation}
\mathbf{x_T} = m_{\mathrm{T}} \odot \mathbf{x}=m_{\mathrm{sem}} \odot m_{\mathrm{ins}} \odot \mathbf{x}
\label{eq:xm}
\end{equation}
where $\odot$ denotes multiplication by element.

\subsection{Instance Semantic Encoding (ISE)}
\label{subsec:ise}

This module encodes the task-masked image $\mathbf{x_T}$ produced by TOIFM into channel symbols under a controllable bandwidth budget. Following the NTSCC paradigm, ISE first extracts a compact latent representation and an optional hyperprior for entropy modeling, and then performs deep JSCC transmission in the latent space.

Specifically, a nonlinear analysis transform maps $\mathbf{x_T}$ to a latent tensor $\mathbf{l}$, and a hyper-encoder produces a hyperprior $\mathbf{p}$ that summarizes the spatially varying statistics of $\mathbf{l}$.
Conditioned on $\mathbf{p}$, a learned entropy model provides entropy estimates for latent elements, which are used to guide adaptive bandwidth allocation.
Finally, a deep JSCC encoder maps $\mathbf{l}$ to channel symbols $\mathbf{u}$ under the bandwidth budget controlled by $\eta$, followed by standard power normalization before transmission.
The hyperprior $\mathbf{p}$ can be optionally transmitted as side information, or used only at the transmitter for rate allocation.

\subsection{Instance Semantic Decoding (ISD)}
\label{subsec:isd}

This module reconstructs the transmitted content from the received symbols and outputs the recovered task-masked image for downstream tasks.
After channel corruption, the receiver observes the noisy symbol sequence $\hat{\mathbf{u}}$ and estimates the latent representation before synthesizing the reconstruction.

When the hyperprior is transmitted, the receiver first recovers $\hat{\mathbf{p}}$ and derives the corresponding latent statistics to refine latent recovery, consistent with hyperprior-aided decoding in NTSCC.
Otherwise, ISD estimates the latent using only $\hat{\mathbf{u}}$, trading some reconstruction fidelity for lower overhead.
The final output is obtained by a nonlinear synthesis transform, yielding $\hat{\mathbf{x}}_T$, which is then fed to subsequent task modules.

\section{Performance Evaluation}

\subsection{Simulation Settings}

We evaluate InsCom on the Visual Genome (VG) and Cityscapes datasets. VG contains 108{,}077 Internet images with rich object and relation annotations, while Cityscapes provides 24{,}998 urban street images with fine and coarse pixel-level labels, representing typical ICV environments. All images are resized to a fixed resolution, and task-critical masks are generated according to the instance-level criteria described in Section~\ref{subsec:toifm}. The wireless link is modeled as an additive white Gaussian noise (AWGN) channel.

We compare four schemes: 1) \textbf{BPG+LDPC}, a conventional separate source--channel design that combines BPG image compression with LDPC codes; 2) \textbf{DeepJSCC}, an autoencoder-based joint source--channel coding scheme; 3) \textbf{NTSCC}, an advanced DeepJSCC variant with nonlinear transforms and variable-rate coding; and 4) \textbf{InsCom}, the proposed instance-aware semantic communication system built on top of NTSCC with additional instance differentiation and task-oriented filtering.

For image transmission quality, we adopt PSNR as a standard metric. To better capture the fidelity in task-critical regions, inspired by ROI-based semantic metrics, we further use Task-Critical PSNR (TC-PSNR). Let $M_{i,j}\in\{0,1\}$ denote the binary mask of task-critical pixels, then the mean squared error in these regions is
\begin{equation}
    \mathrm{MSE}_{\mathrm{TC}} = 
    \frac{1}{\sum_{i=1}^{m} \sum_{j=1}^{n} M_{i,j}}
    \sum_{i=1}^{m} \sum_{j=1}^{n} M_{i,j}
    \big(P_{i,j} - \hat{P}_{i,j}\big)^2,
\end{equation}
and TC-PSNR is computed as
\begin{equation}
    \mathrm{TC\!-\!PSNR} =
    10 \log_{10}\!\left(
    \frac{\mathrm{MAX}^2}{\mathrm{MSE}_{\mathrm{TC}}}
    \right),
\end{equation}
where $\mathrm{MAX}=255$ for 8-bit images. Higher TC-PSNR indicates better reconstruction in task-critical regions.
\begin{figure*}[t]
\centering
\setlength{\tabcolsep}{3pt}     
\renewcommand{\arraystretch}{0.15} 
\resizebox{\textwidth}{!}{%
\begin{tabular}{@{}m{0.115\textwidth} *{3}{m{0.115\textwidth}} m{0.9cm} *{3}{m{0.115\textwidth}}@{}}

\multicolumn{1}{c}{} &
\multicolumn{1}{c}{\footnotesize\textbf{-1 dB}} &
\multicolumn{1}{c}{\footnotesize\textbf{2 dB}} &
\multicolumn{1}{c}{\footnotesize\textbf{5 dB}} &
\multicolumn{1}{c}{} &
\multicolumn{1}{c}{\footnotesize\textbf{-1 dB}} &
\multicolumn{1}{c}{\footnotesize\textbf{2 dB}} &
\multicolumn{1}{c}{\footnotesize\textbf{5 dB}} \\
\noalign{\vspace{2pt}}

\includegraphics[width=\linewidth]{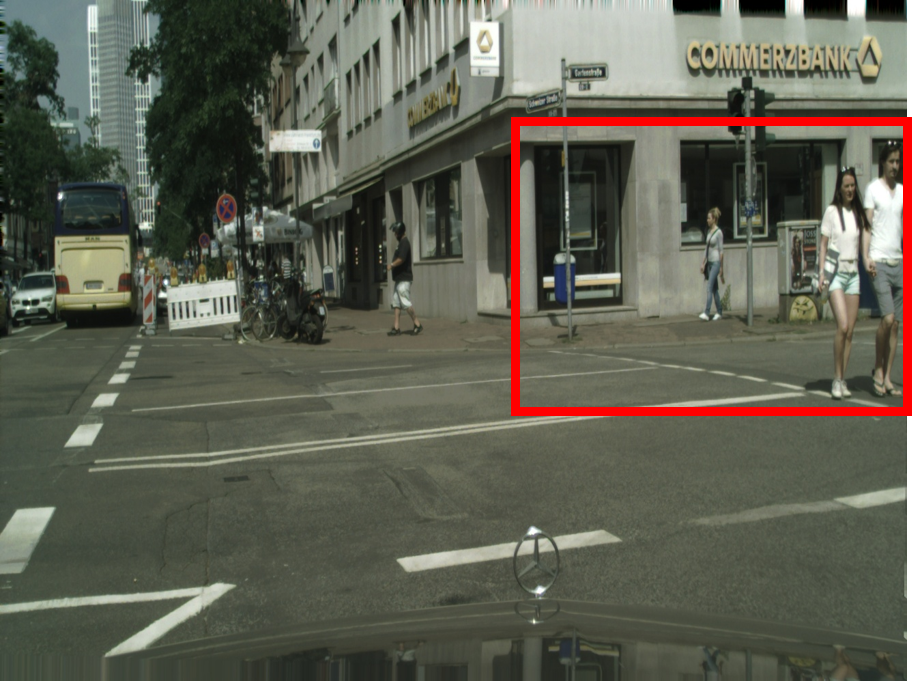} &
\includegraphics[width=\linewidth]{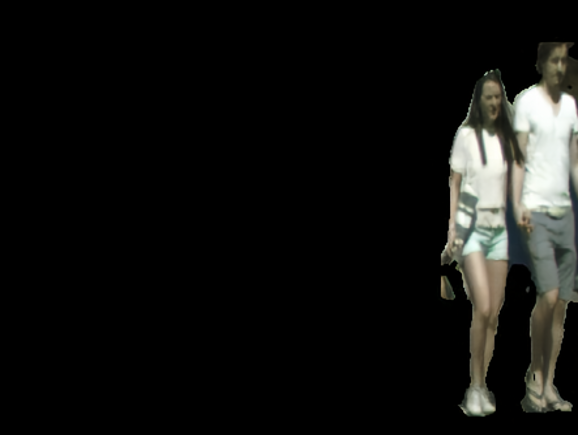} &
\includegraphics[width=\linewidth]{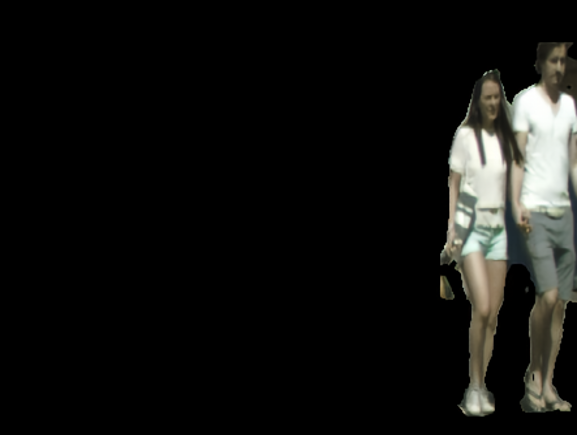} &
\includegraphics[width=\linewidth]{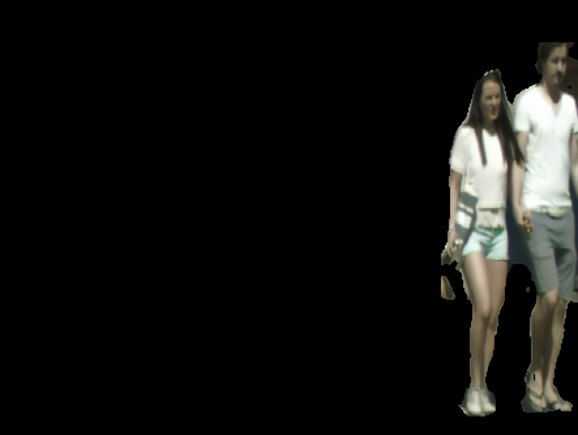} &
&
\includegraphics[width=\linewidth]{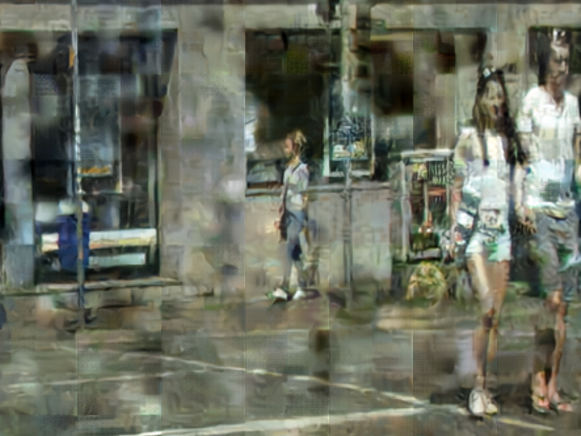} &
\includegraphics[width=\linewidth]{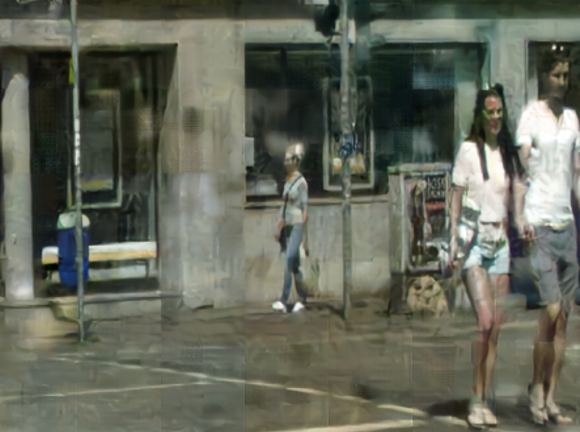} &
\includegraphics[width=\linewidth]{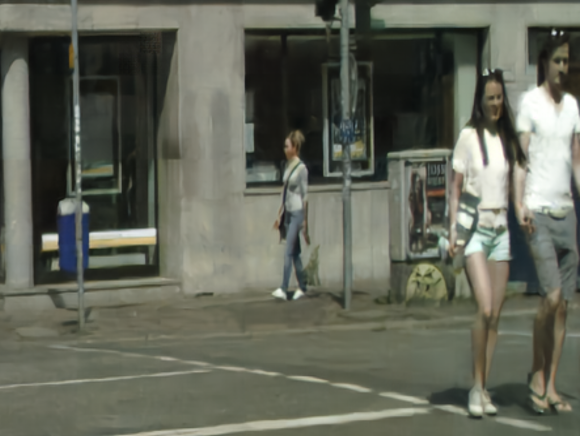} \\
\noalign{\vspace{2pt}}  

\multicolumn{1}{c}{\footnotesize\textbf{Source frame}} &
\multicolumn{3}{c}{\footnotesize (a) InsCom} &
&
\multicolumn{3}{c}{\footnotesize (b) NTSCC} \\
\noalign{\vspace{2pt}}  

\multicolumn{1}{c}{} &
\includegraphics[width=\linewidth]{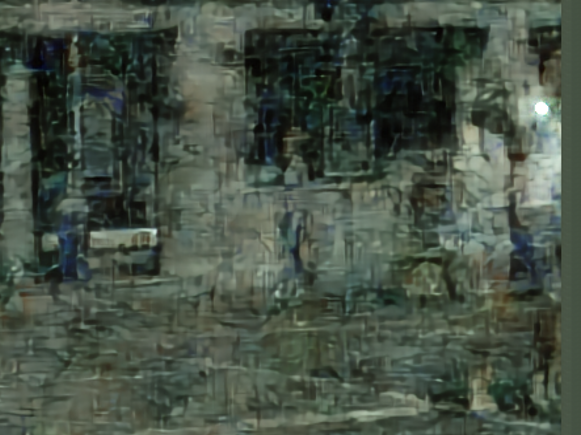} &
\includegraphics[width=\linewidth]{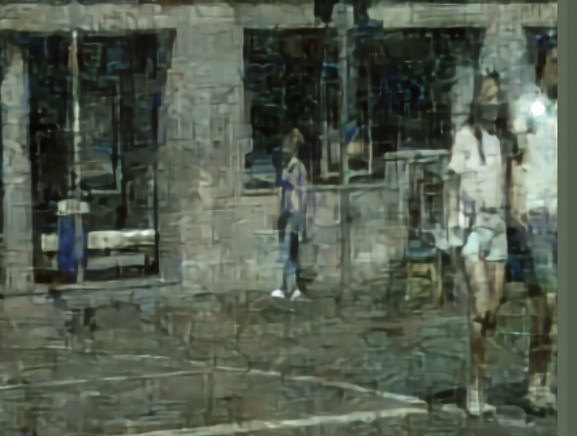} &
\includegraphics[width=\linewidth]{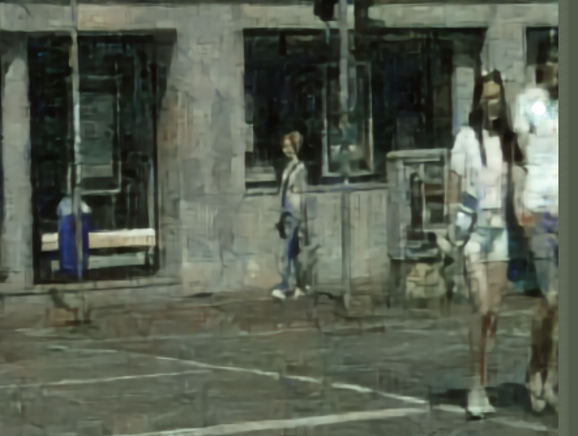} &
&
\includegraphics[width=\linewidth]{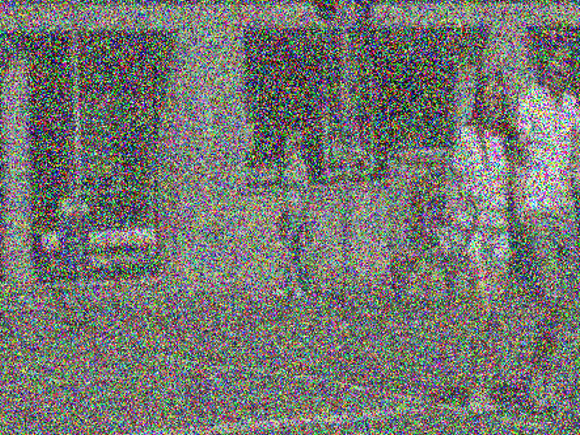} &
\includegraphics[width=\linewidth]{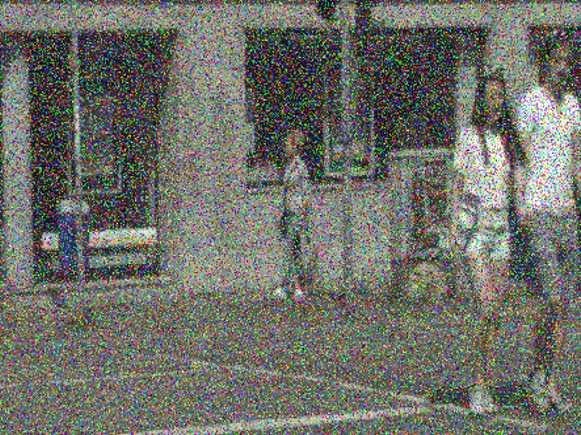} &
\includegraphics[width=\linewidth]{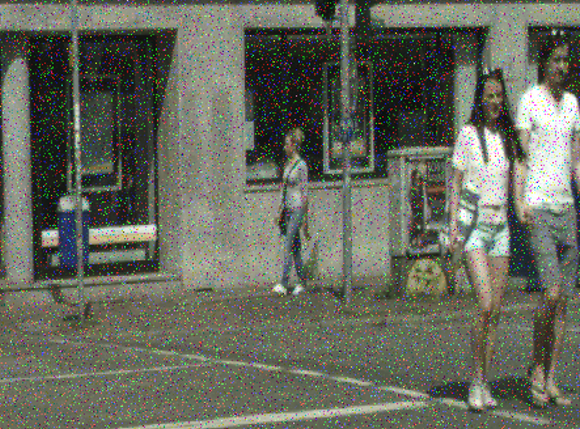} \\
\noalign{\vspace{2pt}}  

\multicolumn{1}{c}{} &
\multicolumn{3}{c}{\footnotesize (c) DeepJSCC} &
&
\multicolumn{3}{c}{\footnotesize (d) BPG+LDPC} \\

\end{tabular}%
}
\vspace{-1mm}
\caption{Visual comparison of TC regions under SNR $=-1/2/5$ dB.}
\label{visual_snr}
\vspace{-2mm}
\end{figure*}

\subsection{Visual Comparison of Reconstruction Performance}

Fig.~\ref{visual_snr} illustrates the reconstruction performance of four schemes for the pedestrian collision warning task within task-critical (TC) regions across varying signal-to-noise ratios (SNR).\par

From Fig.~\ref{visual_snr}, we can find that 1) Reconstruction quality improves with increasing SNR for all schemes. When SNR falls below 0 dB, BPG+LDPC suffers a significant performance drop due to LDPC decoding failures, whereas JSCC maintains better quality through end-to-end nonlinear mapping that preserves semantic features. 2) NTSCC dynamically allocates bit-rates using a variational entropy model, focusing on high-frequency semantic features like pedestrians and buildings more than JSCC's uniform approach, thus performing better in these areas. However, it experiences significant shape and texture distortion at very low SNRs (e.g., -1 dB). 3) InsCom enhances NTSCC by isolating and refining regions of interest, distinguishing between TC-instances within the 'people' category (e.g., prioritizing people on the street over those on sidewalks). Under identical resource constraints, InsCom allocates more resources to each critical instance, achieving superior reconstruction for TC instances (such as walking couples), reducing distortion, and preserving fine-grained shapes and textures.\par
\begin{figure*}[t]
\centering
\setlength{\tabcolsep}{2pt}
\renewcommand{\arraystretch}{0}

\begin{tabular}{@{}ccc@{}}
\scriptsize\textbf{SNR = $-3$ dB} &
\scriptsize\textbf{SNR = $0$ dB}  &
\scriptsize\textbf{SNR = $3$ dB}  \\[0pt]

\includegraphics[width=0.33\textwidth]{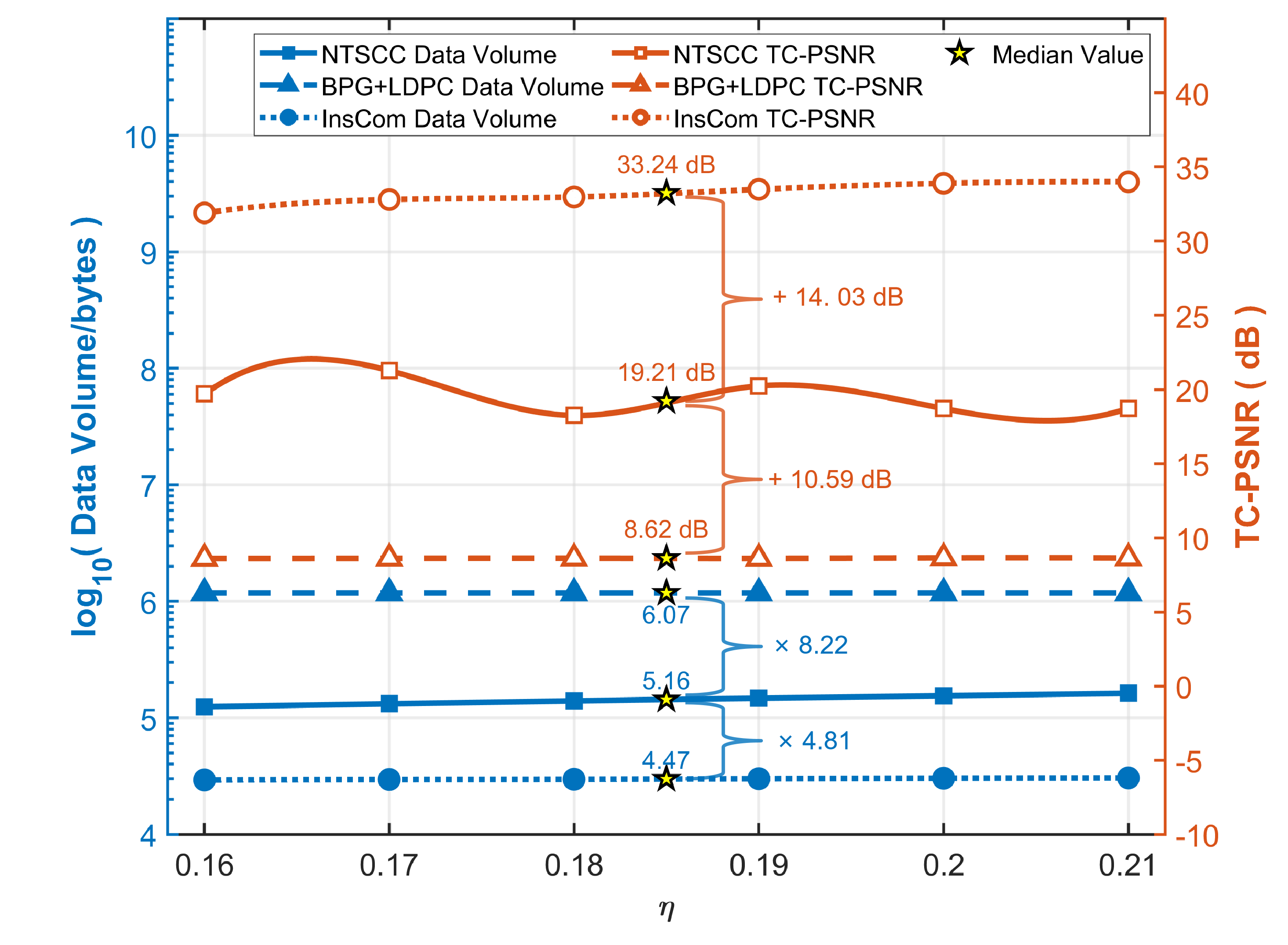} &
\includegraphics[width=0.33\textwidth]{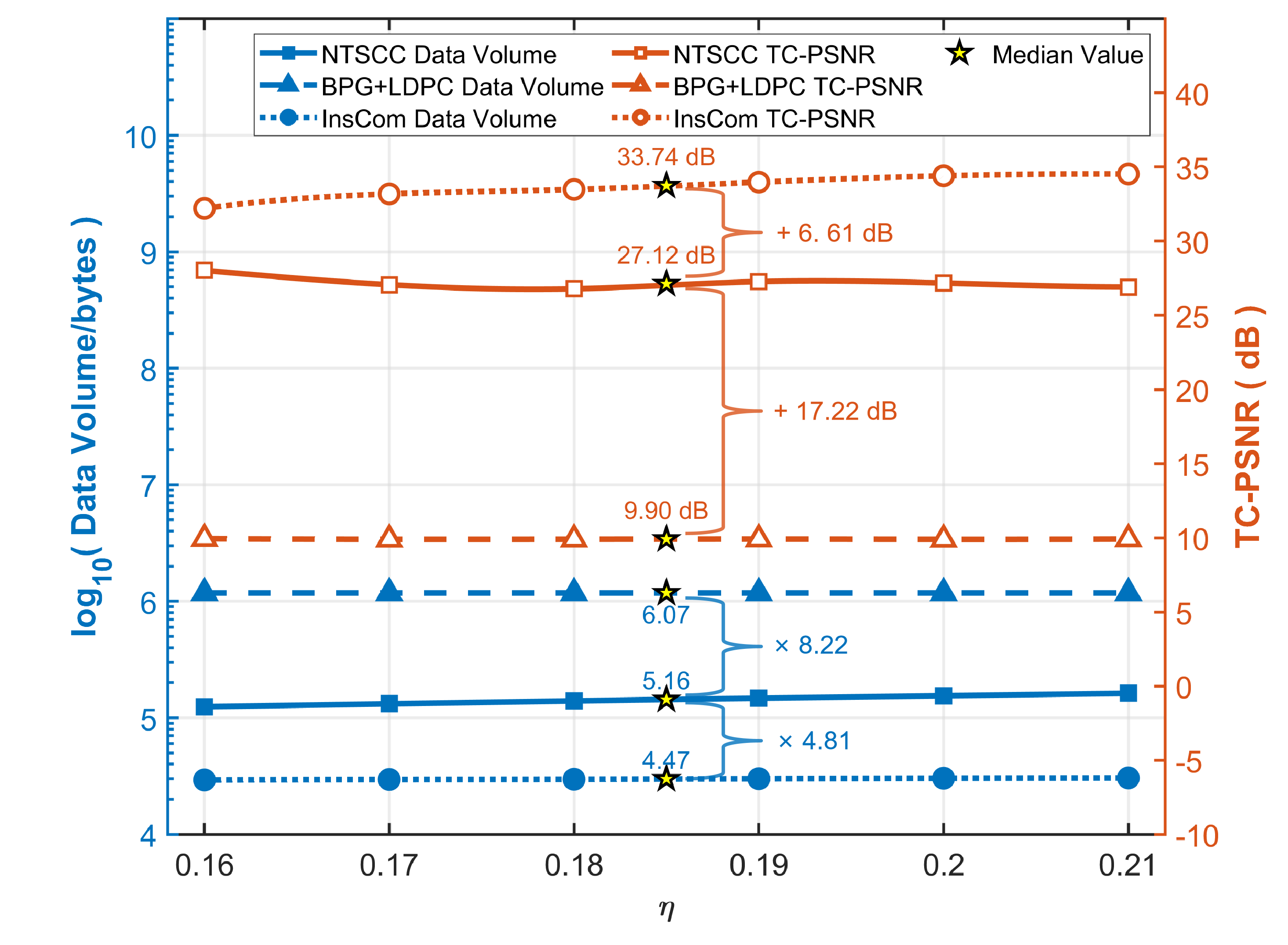}  &
\includegraphics[width=0.33\textwidth]{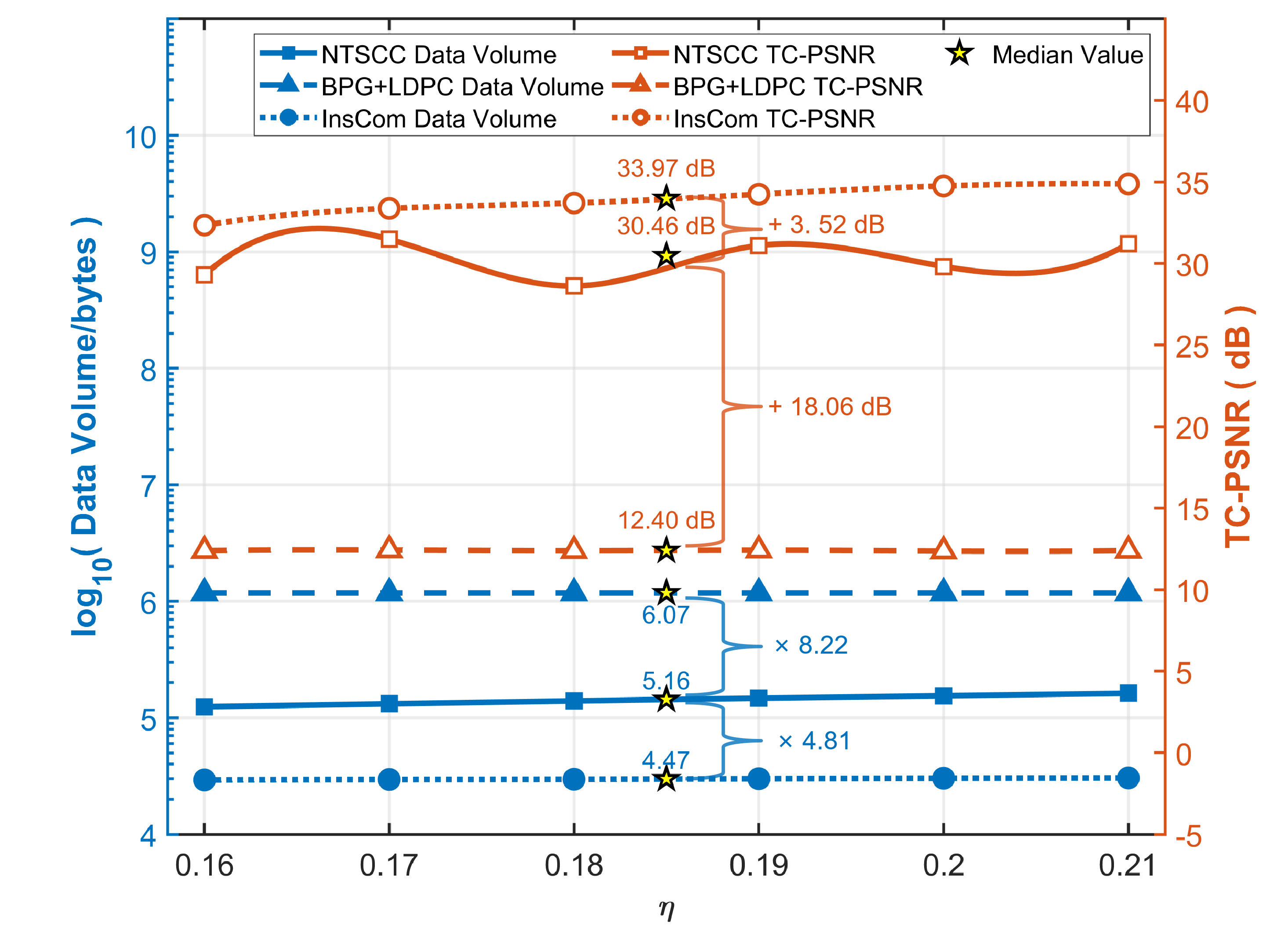}  \\[1pt]

\includegraphics[width=0.33\textwidth]{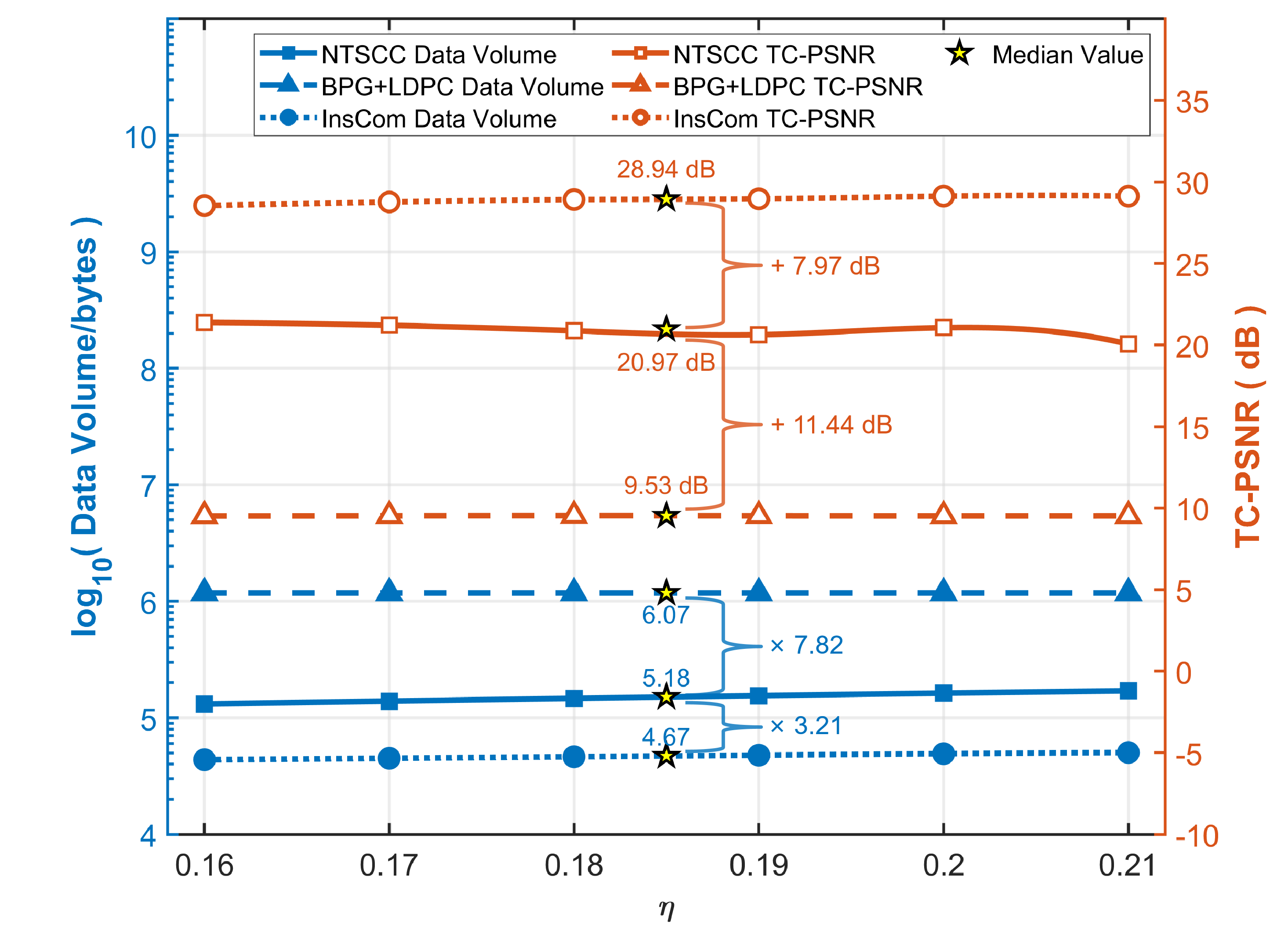} &
\includegraphics[width=0.33\textwidth]{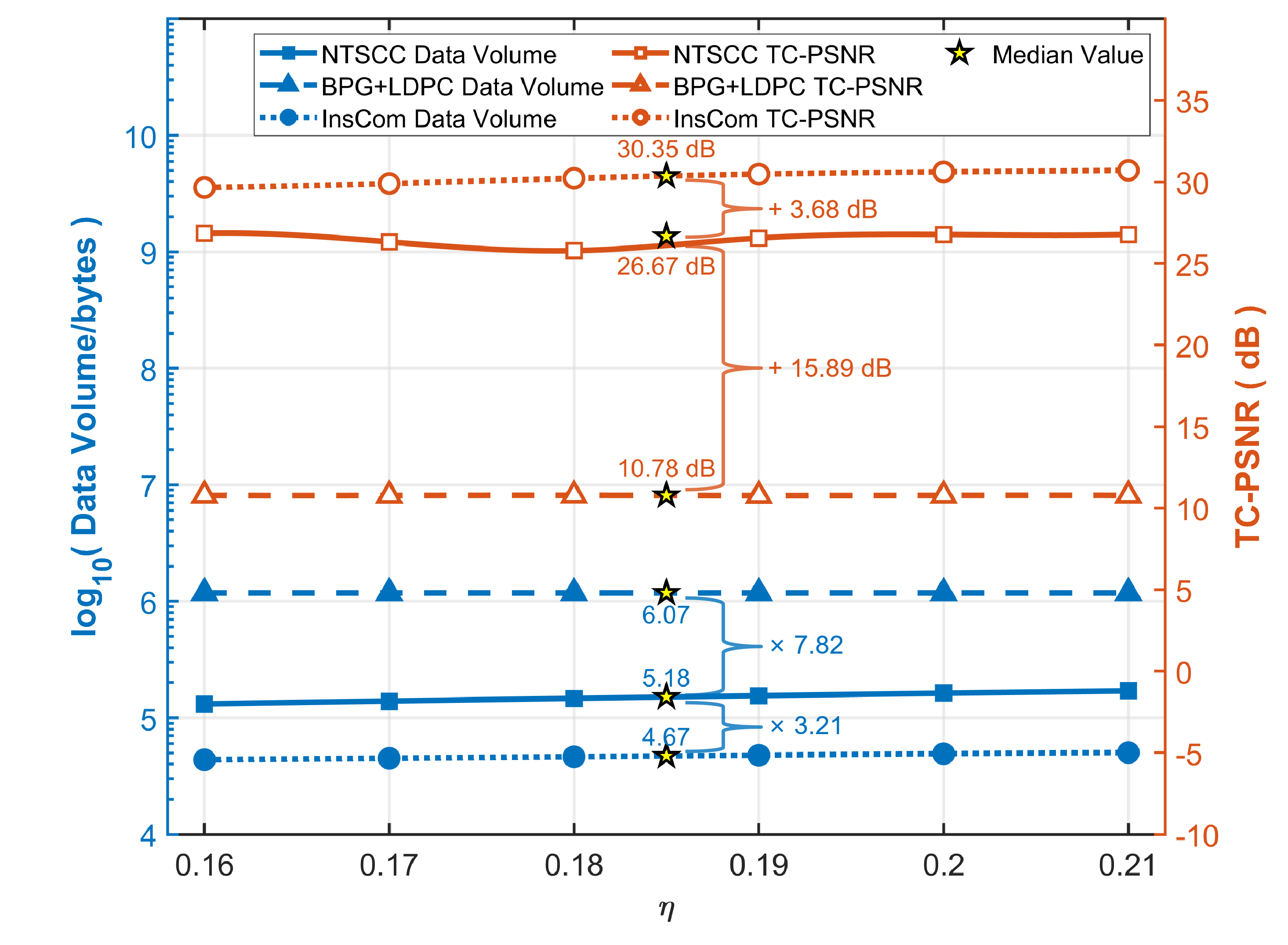}  &
\includegraphics[width=0.33\textwidth]{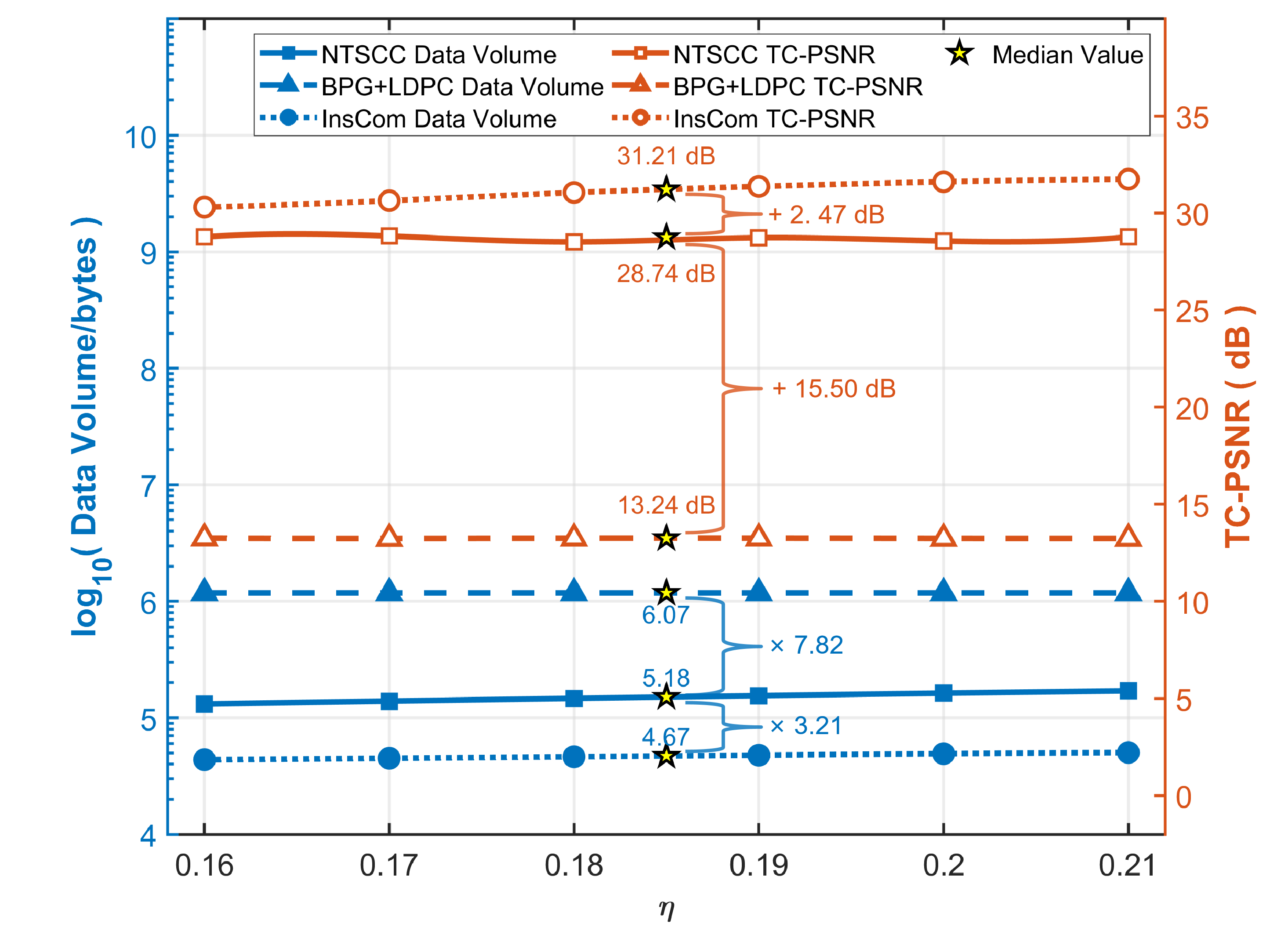}  \\
\end{tabular}

\vspace{-2mm}
\caption{Relationship between required data transmission volume and transmission quality for different schemes across various SNR levels and datasets. First row: Visual Genome dataset. Second row: Cityscapes dataset.}
\label{fig:tc_metrics}
\vspace{-2mm}
\end{figure*}

\subsection{Comparison of Rate-Distortion Trade-Off Performance}

To jointly assess transmission efficiency and reconstruction quality, we sweep the rate–distortion control parameter $\eta$ in both NTSCC and InsCom. Specifically, increasing $\eta$ prioritizes reconstruction quality at the expense of higher data rates by shifting semantic information towards higher-rate quantization intervals. This parametric sweep across $\eta$ enables a thorough evaluation of the rate-distortion trade-off under varying channel conditions. Comparative analyses are conducted at three SNR levels ($-3$, $0$, and $3$dB) on both the VG and Cityscapes datasets.\par

As observed in Fig.~\ref{fig:tc_metrics}, we can find that: 1) Increasing $\eta$ yields gradual \textit{TC-PSNR} improvements with only minimal variations in data volume across all evaluated methods, where InsCom demonstrates the most significant gains. For instance, InsCom achieves a steady increase from 33.24 to 34.10 dB at 3 dB SNR on the VG dataset. 2) Median performance points (denoted by star markers) reveal that NTSCC reduces data volume by 3.21$\times$ while improving quality ranging from 10.59 to 18.06 dB compared to BPG+LDPC systems, thereby confirming the advantage of semantic-level communication. 3) Building on this, InsCom further reduces the data volume by 7.82 times and achieves quality gains ranging from 1.75 to 14.03 dB compared to NTSCC under identical conditions, thereby establishing the clear superiority of instance-level communication.\par 

The performance advantages of InsCom stem from two novel mechanisms: (1) filtering out task-irrelevant instances through semantic and instance-level masks, which effectively reduces source entropy; (2) intelligent bit-prioritization for task-relevant latent features, enabled by the ISE entropy model \( P(\mathbf{y}|\mathbf{z}) \), where \(\eta\) optimally increases to refine this allocation without causing bitstream inflation. This approach stands in direct contrast to NTSCC's uniform bit distribution across identical semantic regions. Despite utilizing the same underlying entropy modeling framework as InsCom, NTSCC leads to inefficient compression.

\section{Conclusion}
This paper introduces InsCom, a system that advances SemCom from the semantic to the instance level. It leverages scene-graph generation to identify and differentiate between instances sharing the same semantics, followed by configurable instance filtering, to effectively reduce transmission redundancy. Extensive evaluation across diverse datasets and wireless channels demonstrates that InsCom achieves over 7.82× lower data volume and a quality gain of 1.75–14.03 dB compared to state-of-the-art SemCom systems. Although the instance differentiation and filtering modules introduce additional computational overhead (130.8 GFLOPs), this cost remains substantially lower than that of the NTSCC backbone (774.0 GFLOPs). Future work will focus on lightweight model designs to reduce latency while maintaining InsCom’s transmission performance gains.



\bibliographystyle{IEEEtran}
\bibliography{references}


\vfill

\vfill

\end{document}